\documentclass[12pt]{article}

\newcommand{\bbr}{I\!\! R}
\newcommand{\bbz}{Z\!\!\! Z}

\newcommand{\3}{$^3$}

\newcommand{\x}{arXiv:}
\newcommand{\m}{\mathrm}

\usepackage{graphicx}

\begin{document}
\thispagestyle{empty}
\begin{center}

\null \vskip-1truecm \vskip2truecm

{\Large{\bf \textsf{Arrow of Time in String Theory} }}

\vskip1truecm

{\large \textsf{Brett McInnes}}

\vskip1truecm

\textsf{\\  National
  University of Singapore}

\textsf{email: matmcinn@nus.edu.sg}\\

\end{center}
\vskip1truecm \centerline{\textsf{ABSTRACT}} \baselineskip=15pt
\medskip
Inflation allows the problem of the Arrow of time to be understood
as a question about the structure of spacetime: why was the
\emph{intrinsic} curvature of the earliest spatial sections so
much better behaved than it might have been? This is really just
the complement of a more familiar problem: what mechanism prevents
the \emph{extrinsic} curvature of the earliest spatial sections
from diverging, as classical General Relativity suggests? We argue
that the stringy version of ``creation from nothing", sketched by
Ooguri, Vafa, and Verlinde, solves both of these problems at once.
The argument, while very simple, hinges on some of the deepest
theorems in global differential geometry. These results imply that
when a spatially toral spacetime is created from nothing, the
earliest spatial sections are forced to be [quasi-classically]
exactly locally isotropic. This local isotropy, in turn, forces
the inflaton into its minimal-entropy state. The theory explains
why the Arrow does \emph{not} reverse in black holes or in a
cosmic contraction, if any.

\newpage

\addtocounter{section}{1}
\section* {\large{\textsf{1. The Arrow of Time: What Inflation Does For Us}}}
One of the deepest mysteries in physics is the origin of the Arrow
of time. The first step towards understanding this mystery was
identifying where the explanation is to be looked for: in
\emph{cosmology}. For example, the immediate origin of the
low-entropy conditions in our local environment is the Sun; and the
Sun's ability to play this role can readily be traced back to the
extremely low total entropy of the Big Bang era. [Excellent surveys
of this question have been given by Albrecht \cite{kn:albrecht},
Price \cite{kn:price}, and Carroll and Chen \cite{kn:carroll}; see
also \cite{kn:coule} for relevant general background.] The problem
is now in the cosmological domain: why was the entropy so low at
that time ---$\,$ what set up the conditions that allowed the
nucleosynthesis which provided the raw materials for the Sun,
together with the subsequent local contraction which formed it?

It is important to understand that we have no right to expect a
single answer to this question. The low entropy of the very earliest
universe might well have been stored in a vast variety of ways, as
Price \cite{kn:price} emphasises. In fact, however, the theory of
Inflation \cite{kn:dolgov} provides a beautifully simple answer. The
matter and radiation of the early Big Bang era derived its low
entropy from a single source: the inflaton. For it was the inflaton
that generated the equilibrium that obtained just before re-heating,
and it was of course the inflaton that drove re-heating itself. This
is the first service that Inflation performs in our search for the
origin of the Arrow of time: it reduces what could have been an
enormously complex collection of problems [explaining a whole
variety of ways in which the initial low entropy state might have
been arranged] to a single one: why was the inflaton itself in a
special state \cite{kn:trodden} in the beginning?

The special initial state of the inflaton had a precise form: the
inflaton was in a ``potential-dominated" state. The
stress-energy-momentum tensor for the inflaton has the form
\begin{equation}\label{eq:A}
\m{T_{\mu\nu}\;=\;\partial_{\mu}\varphi\;\partial_{\nu}\varphi\;-\;
{{1}\over{2}}\,g_{\mu\nu}\,[g^{\alpha\beta}\,\partial_{\alpha}\varphi\;\partial_{\beta}\varphi\;-\;V(\varphi)].}
\end{equation}
This expression is dominated by the potential V($\varphi$), allowing
the inflaton to mimic a cosmological constant, provided that this
potential is much larger than the kinetic term, involving the
\emph{derivatives} of the inflaton field. Among all of the vast
array of possible excitations of the degrees of freedom of
$\varphi$, only an infinitesimal fraction would have satisfied this
condition; thus, ``low entropy" here means something very specific:
that the \emph{gradient vector} of the inflaton field was
essentially zero at the earliest times.

The question as to whether Inflation itself can explain this special
initial state of the inflaton has given rise to much debate
\cite{kn:mukhanov}\cite{kn:hollands}. Recently, however, it has
become generally agreed that string theory ---$\,$ which will be the
context for this work ---$\,$ always maintains strict
\emph{unitarity}, even under extreme circumstances
\cite{kn:juan}\cite{kn:stephen}\cite{kn:sbound}. This suggests [see
\cite{kn:carroll} for the details of the argument] that Inflation
\emph{alone} cannot solve the problem of the Arrow of time. Albrecht
\cite{kn:albrecht} has given more general reasons in favour of this
conclusion. But if Inflation itself is not the solution, it does
firmly point towards the direction in which a full account must be
sought, as follows.

The question of the origin of the Arrow of time has now been reduced
to this question: why was the gradient vector of the inflaton so
small in the beginning? The most natural way to explain why a
\emph{spatial} vector field should be zero is to invoke \emph{local
isotropy} at each point of space\footnote{\emph{Local} isotropy at a
point in a Riemannian manifold is the condition that, given a point
p and a pair of unit tangent vectors X, Y, at p, there exists an
open neighbourhood of p and an isometry of this neighbourhood which
maps X to Y. Note that even if the manifold is everywhere locally
isotropic, it \emph{need not} be globally isotropic.}. This is based
on the simple observation that a vector can be isotropic only if it
vanishes.

We propose, then, that local spatial isotropy was the specific
geometric property which enforced the vanishing of the initial
inflaton gradient. Note that for this argument to work, the isotropy
must be very precise ---$\,$ we are not speaking of the usual,
approximate isotropy discussed in connection with observational
cosmology. Instead, we need a fundamental isotropy which, in fact,
is \emph{exact} at the quasi-classical level. This, again, is an
extremely ``special" state for the spatial geometry at \emph{any}
time, and, in view of the way anisotropies tend to grow when one
traces the history of a spatial section back in time, it is still
more so at early times. Leaving aside [for the moment] the question
as to why the temporal gradient of the inflaton should have been
initially zero, we see that the question has been transferred to the
domain of spatial geometry: \emph{why were the earliest spatial
sections exactly locally isotropic} at the quasi-classical level?

The idea that the Arrow of time is ultimately due to the uniformity
of the initial spacetime geometry is due to Penrose
\cite{kn:penrose}. Penrose, too, implicitly emphasises the role of
local isotropy, by focusing on the Weyl curvature, which is a
measure of tidal anisotropy. In our view, Inflation is an extremely
natural idea in this context, because a scalar field, unlike the
vector or spinor fields which represent all of the known fundamental
forms of matter, can be non-trivial even on a perfectly locally
isotropic spacetime. An initial geometry of this kind will
automatically eliminate all vectors and spinors, and it will
automatically put a scalar field into the potential-dominated state.
Thus, initial local isotropy explains [a] why a \emph{scalar} field
dominated other possible forms of matter in the earliest Universe,
and [b] why that scalar field was in a particular, low-entropy state
at that time. That is, Inflation provides an extremely natural way
of communicating the initial \emph{geometric} ``specialness" to all
other physical fields.

To summarize, then: Inflation itself does not explain the Arrow of
time, but it provides a major part of the explanation. First, it
enormously simplifies the problem, by reducing it to explaining the
low initial entropy of a \emph{single} object, the inflaton; and
second, it allows us to reduce this problem to explaining the local
isotropy of the earliest spacetime geometry. Conversely, a physical
explanation of the initial local isotropy may well explain why the
inflaton was the dominant form of matter in the beginning.

In \emph{three} dimensions, a vector is dual to a two-form, so a
locally isotropic three-dimensional Riemannian manifold has the
property that its sectional curvature at each point is independent
of direction. By Schur's theorem [see \cite{kn:kobayashi} page 202],
this means that the space has the geometry of a \emph{space of
constant curvature} throughout the [connected] region in which it is
locally isotropic. From the spacetime point of view, we can say that
the geometry near the beginning had the property that the
\emph{intrinsic curvature} tensor of the spacelike slices was
maximally well-behaved, in the sense that it has the structure of
the curvature tensor\footnote{It may be helpful, in the case of FRW
metrics, to think in terms of the fundamental, (1,3) version of the
intrinsic curvature tensor, which is not affected by the variations
of the scale factor.} of one of the classical spaces of constant
curvature: the sphere, Euclidean space, hyperbolic space, and their
non-singular quotients\footnote{From footnote 1, it is clear that
all of these quotients [such as $\bbr$P$^3$ and the torus T$^3$] are
indeed \emph{locally} isotropic at every point.}. In order to
explain the Arrow of time, we now have a specific task: to explain
this good behaviour of the intrinsic curvature of the earliest
spatial slices.

Now, in fact, it is generally believed that the \emph{extrinsic}
curvature of the earliest spatial sections was \emph{also} much
better behaved than one has a right to expect. That is, classical
General Relativity leads one to expect that the extrinsic curvature
should diverge as earlier times are examined, but it is generally
believed that string theory intervenes to prevent this from
happening. In fact, string theory has given rise to a variety of
ideas as to how the misbehaviour of the extrinsic curvature might be
removed or otherwise tamed. One recent proposal for using the theory
to deal with cosmological singularities was made by Ooguri et al
\cite{kn:ooguri}\cite{kn:OVV}, who have embedded the idea of
``creation from nothing" in string theory; another suggestion, due
to McGreevy and Silverstein \cite{kn:end}, involves a
pre-inflationary era of tachyon condensation.

Since the mystery of the Arrow of time is just the ``intrinsic
curvature version" of the ``extrinsic curvature problem" ---$\,$ the
problem of resolving cosmological singularities ---$\,$ it is
natural to demand that any theory that claims to solve one of these
problems should solve the other. Indeed, it is clearly pointless to
resolve cosmological singularities if the resulting, non-singular
spacetime fails to evolve to a Universe like ours. But that will
surely be the case unless the theory strictly \emph{demands} initial
local spatial isotropy at each point. Equally, it seems very
unlikely that a geometric solution of the problem of the Arrow of
time can ignore the singularity problem. In other words,
\emph{string theory must produce an initial spacetime structure
which is extremely uniform not only in the timelike, but also in the
spacelike directions:} it must account for the good behaviour of
extrinsic \emph{and} intrinsic curvature. This is, of course, an
extremely formidable task.

Here we shall argue that the version of string cosmology due to
Ooguri et al \cite{kn:ooguri} may in fact be capable of this feat.
In this theory, the Universe is ``created from nothing", after the
manner of Vilenkin \cite{kn:vilenkin} or Hartle and Hawking
\cite{kn:hartle} [though the details are very different]. ``Creation
from nothing" fits particularly well into our discussion above,
because it takes place along a hypersurface of \emph{time symmetry};
along such a hypersurface, by definition, the extrinsic curvature
vanishes, so the problem of a divergent extrinsic curvature is
solved by the same mechanism that allows ``creation from nothing" in
the first place. Furthermore, creation along a hypersurface of time
symmetry immediately explains why the initial \emph{temporal}
gradient of the inflaton should vanish, something not ensured by
local spatial isotropy.

While this proposal automatically solves the extrinsic curvature
problem, there is no reason to expect the initial intrinsic
curvature to be well-behaved: the initial spatial section could in
principle have had an extremely distorted non-singular geometry, and
indeed that would have been the generic case. The task now is to
show that the wave function of Ooguri et al \cite{kn:ooguri} is
extremely sharply peaked around some three-dimensional geometry of
constant curvature.

Unfortunately, the ``OVV wave function" is not understood in enough
detail to check this \emph{directly}. This difficulty can, however,
be circumvented by appealing to topological methods. We shall see
that a simple argument ---$\,$ based, however, on very deep theorems
due to Schoen, Yau, Gromov, Lawson, and Bourguignon ---$\,$ shows
that, in fact, a spatial section can only be ``created from nothing"
in the OVV framework if it is, classically, \emph{exactly a space of
constant curvature}: in fact, classically it must be exactly locally
flat. This is the origin of the extreme initial geometric
``specialness" which, via the inflaton, gives rise to the Arrow of
Time.

The Ooguri et al version of string cosmology is in its infancy, and
there are many subtle aspects of this argument which remain to be
clarified. It is nevertheless of interest to understand one possible
way in which string theory might account for the Arrow of time. More
generally, it is of interest to exhibit a theory in which the Arrow
of time is explained \emph{without} resorting to vastly improbable
random fluctuations, to multiple universes, or to the anthropic
principle.

We begin with a brief review of the relevant aspects of the theory
of Ooguri et al. We then discuss the particular form which the
initial value problem takes in the case of
spacetimes-with-boundaries. Finally, we give the [logically
extremely simple, but mathematically very deep] argument that OVV
cosmology implies an Arrow, and discuss its consequences. We
specifically address Price's \cite{kn:price} concern that
beginnings and endings [particularly inside black holes] should be
treated symmetrically.

The Arrow of time problem has recently attracted increased
attention, giving rise to a variety of interesting ideas: see for
example \cite{kn:carroll}\cite{kn:laura}\cite{kn:gorsky}. In our
view, it is not enough to explain why the entropy of the early
Universe was low: we must explain why it was low \emph{in a
particular way}, namely, in the form of maximally well-behaved
intrinsic curvature. Furthermore, for reasons already explained, we
believe that an adequate explanation of the Arrow of time can only
arise in connection with an explicit solution of the cosmological
singularity problem. For this reason, we shall be concerned only
with ideas arising directly from string theory, to which we now
turn.

\addtocounter{section}{1}
\section*{\large{\textsf{2. String Theory and The Arrow of Time: General Background}}}
We have argued that the problem of the Arrow of time cannot be
separated from the cosmological singularity problem.  As it is
generally believed that the singularity problem must be solved by
some feature of string theory, that same feature must give rise to
the Arrow of time. In view of this, let us briefly draw attention to
some relevant properties of string cosmology. We begin with some
general observations which might apply to any string-theoretic
account of the Arrow of time, and then focus on the relevant
properties of the theory of Ooguri et al.

\subsubsection*{{\textsf{2.1. String Theory and Spatial Compactness}}}
It is now generally agreed that Inflation is \emph{not} past-eternal
\cite{kn:borde}. The natural interpretation of this is that the
Universe had a beginning, which may or may not have been singular.
Our problem is to understand the properties of the spacelike
hypersurface along which the Universe came into existence. In
particular, we need to answer a basic question: is space, like
[past] time, necessarily finite?

Studies of the earliest Universe in string theory suggest that the
spatial sections of our Universe were, at least in that era,
topologically compact. [We say, ``in that era", because some have
argued that a spatially compact spacetime can, in principle,
ultimately generate spatially non-compact ``babies", as in
\cite{kn:susskind}.] For example, in \emph{string gas cosmology}
\cite{kn:bat}, T-duality is applied to the spatial sections of the
Universe, which are assumed to have the topology of a
three-dimensional torus. In fact, such spatial topologies are
natural in any cosmology involving closed strings; for example,
windings of closed strings are crucial in the string tachyon
cosmology of McGreevy and Silverstein \cite{kn:end}. Finally,
spatial compactness plays an essential role in the ``creation from
nothing" approach, which has been embedded in string theory by
Ooguri et al. Indeed, Ooguri et al \cite{kn:ooguri} are very
explicit in this regard: they state that one of the principal
objectives of their work is to construct a Hartle-Hawking wave
function which determines the [finite] sizes of \emph{all} of the
spatial dimensions, small and large: the sizes are to be regarded as
``moduli". In this approach to string cosmology, the compactness of
the early spatial sections is fundamental.

To summarize: if we wish to embed cosmology in string theory, we
should assume that the initial spatial sections were compact, though
\emph{not} necessarily with spherical topology.

\subsubsection*{{\textsf{2.2. Background on OVV: Creation on a Torus}}}
With this preparation, let us turn to the relevant details of the
theory of Ooguri et al [henceforth, OVV], which we propose to use
here.

It was suggested in \cite{kn:osv} [see \cite{kn:xi}\cite{kn:pioline}
for recent developments] that the [modified] elliptic genus of a
certain IIB Calabi-Yau black hole is squared norm of the topological
string partition function associated with the corresponding
``attractor geometry", the Euclidean space
$(\m{H^2}/\bbz)\,\times\,\m{S^2\,\times\,CY}$. Here H$^2/\bbz$ is a
partially compactified version of two-dimensional hyperbolic space,
S$^2$ is the two-sphere, and CY denotes a six-dimensional Calabi-Yau
space. OVV propose to put this remarkable development to good use by
interpreting the partition function in terms of a Hartle-Hawking
wave function describing the creation of a two-dimensional
cosmological spacetime obtained by taking a Lorentzian version of
H$^2/\bbz$. [This is the ``Entropic Principle".] That is, instead of
beginning, in the usual Hartle-Hawking manner, with a Euclidean
sphere, OVV begin with a \emph{negatively} curved space with
topology $\bbr\,\times\,\m{S}^1$. The metric, with curvature
$-\,1$/L$^2$, is
\begin{equation}\label{eq:B}
g(\m{H}^2/\bbz)_{++}\;=\;\m{K^2\,e^{(2\,\rho/L)}\,d\tau^2\;+\;d\rho^2},
\end{equation}
where $\tau$ is an angular coordinate on a circle with radius K at
$\rho$ = 0; when $\bbr\,\times\,\m{S}^1$ is endowed with this metric
we can call it H$^2$/$\bbz$.

Ooguri et al. argue that the metric $g(\m{H}^2/\bbz)_{++}$ actually
defines \emph{two} Lorentzian metrics. In one way of thinking about
this metric [which OVV call the ``more traditional" interpretation],
$\tau$ is to be regarded as the usual cyclic Euclidean time, and the
$\tau$ translation generator defines a Witten index associated with
the degeneracy of the states of the black hole in the OSV
equivalence. But, in the other, one interprets $\rho$ as Euclidean
time, and then $g(\m{H}^2/\bbz)_{++}$ is regarded as a sort of
Euclidean version of de Sitter spacetime, with flat but compact
spatial sections parametrized by $\tau$. The Lorentzian version is
obtained by complexifying \emph{conformal} time. It is this
``cosmological interpretation" of H$^2$/$\bbz$ that allows OVV to
embed ``creation from nothing" in string theory. Let us see how this
works in the four-dimensional case \cite{kn:OVV}.

We begin with H$^4/\bbz^3$, obtained by compactifying the sections
of H$^4$ when it is foliated by copies of $\bbr^3$ with its
\emph{flat} metric. This gives us the correct four-dimensional
version of $g(\m{H}^2/\bbz)_{++}$: it is the metric on
$\bbr\;\times\;$T$^3$, where T$^3$ is the three-torus, given by
\begin{equation}\label{eq:C}
g(\m{H}^4/\bbz^3; \Phi)_{++++} \;=\; \m{d\Phi^2\;
+\;K^2\,e^{(-\,2\,\Phi/L)}\,[\,d\theta_1^2 \;+\; d\theta_2^2 \;+\;
d\theta_3^2]},
\end{equation}
where the curvature is again $-\,1$/L$^2$, where $\theta_{1,2,3}$
are angular coordinates on a cubic torus with side lengths 2$\pi$K
at $\Phi$ = 0, and where $\Phi$ is a coordinate which runs from
$-\,\infty$ to $+\,\infty$. We now define a ``Euclidean conformal
time" $\eta_-$, with values in ($-\,\infty$, 0), and the metric
becomes
\begin{equation}\label{eq:D}
g(\m{H}^4/\bbz^3;\eta_-)_{++++} \;=\;\m{{{L^2}\over{\eta_-^2}}\,[\,
d\eta_-^2\;+\;d\theta_1^2 \;+\; d\theta_2^2 \;+\; d\theta_3^2]}.
\end{equation}
Complexifying $\eta_-\,\rightarrow\,\pm\m{i}\eta_+$, where $\eta_+$
takes its values in (0, $\infty$), we obtain a spatially flat
version of Lorentzian de Sitter spacetime, with toral sections, in
($+\;-\;-\;-$) signature:
\begin{eqnarray}\label{eq:E}
g(\m{STdS_4})_{+---} & = & \m{{{L^2}\over{\eta_+^2}}\,[\,
d\eta_+^2\;-\;d\theta_1^2 \;-\; d\theta_2^2
\;-\; d\theta_3^2]} \nonumber \\
                     & = &  \m{dt^2\; -\;K^2\,e^{(2\,t/L)}\,[\,d\theta_1^2 \;+\;
d\theta_2^2 \;+\; d\theta_3^2]},
\end{eqnarray}
where t ranges from $-\,\infty$ to $+\,\infty$. This is Spatially
Toral de Sitter spacetime [see \cite{kn:gall}].

Notice that we can obtain the final equation in (\ref{eq:E}) [up to
an overall sign] from (\ref{eq:C}) in a purely formal way by
complexifying both $\Phi$ and L. This formal procedure has the
advantage of quickly yielding the final answer for more complicated
metrics which may be only asymptotically hyperbolic.

In this simple way, we can give concrete form to the intuition that
partially compactified hyperbolic space can be used to define a
reasonable [that is, inflationary] cosmology. ``Creation from
nothing" in the OVV sense can now be interpreted as follows. We
truncate H$^4$/$\bbz^3$ so that $\Phi$ takes values in ($-\,\infty$,
0], while Spatially Toral de Sitter, STdS$_4$, is truncated so that
t takes values in [0, $\infty$). The two spaces are thus both
truncated along a three-dimensional torus consisting of circles of
circumference 2$\pi$K. We now join the two spaces along this torus,
which is the hypersurface where the transition from Euclidean to
Lorentzian geometry ---$\,$ that is, ``creation from nothing"
---$\,$ takes place. We can summarize the whole construction by
\begin{eqnarray}\label{eq:F}
g(\m{H^4}/\bbz^3;\; -\,\infty\,<\,\Phi\,\leq\,0)_{++++}\;
\longrightarrow\;g(\m{STdS_4;\;0\,\leq\, t\,<\,\infty})_{+---},
\end{eqnarray}
where the arrow symbolizes the transition from a Euclidean to a
Lorentzian metric, bearing in mind that the Euclidean-to-Lorentzian
transition occurs along $\Phi$ = t = 0 [$\eta_-$ = $-\,$L/K and
$\eta_+$ = L/K.]

It turns out \cite{kn:OVV} that this procedure, which allows the OVV
wave function to have a cosmological interpretation, \emph{only}
works when the spatial sections have toral topology. This topology
is therefore a prediction of stringy ``creation from nothing".

This is actually self-consistent, if we recall that one of the
avowed objectives of the OVV theory is to \emph{predict} the size of
the Universe at its birth; in string theory, this size should be
given by the string scale. Note that it is only in the toral case
that the initial size is decoupled from the spacetime curvature
scale
---$\,$ there are two independent length scales, L and K,
in all of the metrics discussed above. This useful property of the
spatially toral version of de Sitter spacetime was applied to
quantum cosmology by Zel'dovich and Starobinsky \cite{kn:zelda}, and
by Linde \cite{kn:lindetypical}; see also \cite{kn:tallandthin}.

\subsubsection*{{\textsf{2.3. How Stringy ``Creation From Nothing" Evades The Singularity Theorems}}}
``Creation from nothing" clearly avoids an initial singularity. That
fact does not absolve us of a duty to explain how the various
relevant singularity theorems are circumvented. This will be needed
later, when we study the initial value problem for ``creation from
nothing".

Technically, ``creation from nothing" avoids an initial singularity
by happening along a spacelike hypersurface of \emph{zero extrinsic
curvature.} We have to explain how this can be arranged when the
initial spatial section has a toral topology.

``Creation from nothing" is formally ---$\,$ not physically
---$\,$ related to ``bounce" cosmologies. The spacetimes in the
latter case have a hypersurface of zero extrinsic curvature at the
bounce, so, by deleting the contracting half of the spacetime and
inserting a boundary, one obtains a ``creation from nothing"
spacetime. Conversely, one can always construct a formal bounce
cosmology from the kind of spacetime we are considering here.
\emph{This formal equivalence to a bounce spacetime explains how
``creation from nothing" spacetimes evade the singularity theorem
of Borde et al \cite{kn:borde}}. For this theorem assumes a
preponderance of expansion over contraction up to the present,
which is not the case for a bounce spacetime.

However, for spacetimes with toral spatial sections there is a much
stronger singularity theorem, and this too must be circumvented. It
is well known that FRW bounce cosmologies with flat spatial sections
violate the Null Energy Condition if the Einstein equations hold. In
fact, this statement has been greatly extended by Andersson and
Galloway \cite{kn:andergall}: one of their results essentially
demands a past singularity if [as is the case for any spacetime
which evolves towards an inflationary state] a spacetime is
non-singular and de Sitter-like to the future, provided only that
the \emph{Null Ricci Condition} is satisfied and that the spatial
sections have the \emph{topology} ---$\,$ not necessarily the
canonical geometry ---$\,$ of a torus. [See \cite{kn:singularstable}
for further discussions of the physical implications of this
remarkable theorem.] We stress that the result holds for \emph{any}
choice of metric on the topological torus.

The Null Ricci Condition [NRC] is the requirement that, for all null
vectors k$^{\mu}$, the Ricci tensor should satisfy
\begin{equation}\label{eq:II}
\mathrm{R}_{\mu\nu}\,\mathrm{k}^\mu\,\mathrm{k}^\nu\;\geq\;0.
\end{equation}
The conclusion of this discussion is that \emph{the OVV version of
``creation from nothing" can only work if the NRC is violated} at
some time, presumably near to the creation.

The NRC is equivalent to the Null Energy Condition [NEC] provided
that the Einstein equations hold exactly. The question as to
whether NEC violation is compatible with string theory has
recently been a subject of great interest: see for example
\cite{kn:crem}\cite{kn:ovrut}\cite{kn:cremi}\cite{kn:nimah}. The
emerging consensus is that NEC violation \emph{is} acceptable,
even in the full string theory, though only under restrictive
conditions: for example, the violations of the NEC involved in
black hole evaporation are acceptable, as of course are those
associated with orientifold planes, and Casimir effects
\cite{kn:coule}\cite{kn:god} [particularly relevant to toral
cosmologies] can also be acceptable in some circumstances.

On the other hand, in the earliest Universe we do not expect the
Einstein equations to hold exactly; if the Einstein equations are
indeed modified then it becomes possible to violate the NRC while
preserving the NEC. We can call this \emph{effective} violation of
the NEC. Again, we need to ask whether effective violation of the
NEC is possible in the context of the OVV theory
---$\,$ that is, in IIB string theory.

Fortunately, the answer to this is known: the existence of bouncing
cosmological solutions in this case was shown explicitly by Kachru
and McAllister \cite{kn:kachru}, who investigated a brane cosmology
in the context of the Klebanov-Strassler conifold in IIB theory.
[More generally, the possibility of effective NEC violation in brane
world cosmologies has been clearly explained in \cite{kn:varun}; see
also \cite{kn:ruth}.] This can be regarded as an \emph{effective}
scalar-tensor theory; it is well known that such theories can easily
give rise to effective violations of the NEC. [They are of
considerable interest also with regard to the current accelerated
expansion: see for example
\cite{kn:star}\cite{kn:peri}\cite{kn:nood}.]

Other ways are known of violating the NRC while avoiding the
instabilities that often arise from genuinely NEC-violating matter
fields; for example, classical constraint fields \cite{kn:gab},
metric-affine theories and so on \cite{kn:stach}. One particularly
interesting possibility \cite{kn:where}\cite{kn:biswas} involves
ghost-free higher derivative corrections of the Einstein-Hilbert
Lagrangian. These arise quite naturally in string theory
\cite{kn:ghost}.

In all these cases, including the effective theory considered by
Kachru and McAllister, the NRC-violating effect \emph{can be
formally represented by a ``fluid" with a negative ``energy density"
and a causal equation of state}. [NEC violation is often associated
with causality violation, but that is only the case if the energy
density is positive. For a perfect fluid with negative energy
density $\rho$ and pressure p, there will be no propagation outside
the local light cone, for any local observer, provided that
$\m{|\,p|\;\leq\;|\,\rho|}$; this, however, automatically entails an
effective violation of the NEC. That is, for a fluid with negative
energy density, causality automatically ensures that the NEC is
\emph{violated}, not preserved.]

The equation-of-state parameter need not be constant, but in simple
cases it is: for example, Casimir effects are represented by a
``fluid" with a negative energy density and an equation-of-state
parameter equal to 1/3, while the Gabadadze-Shang constraint field
\cite{kn:gab} can be represented formally as a ``fluid" with
negative ``energy density" and equation-of-state parameter equal to
1. [An alternative way of thinking about these generalized
cosmological models is in terms of an effective [classical]
potential, as in \cite{kn:syd}; essentially all of the many models
considered in the literature can be described in this way.] We shall
assume henceforth that whatever it is that violates the NRC in the
OVV theory ---$\,$ thereby permitting the existence of a boundary of
zero extrinsic curvature
---$\,$ can be represented formally by a ``fluid" [or classical
effective potential].

Let us examine an example of the kind of metric that results when
the inflaton [represented at the creation by a positive cosmological
constant] is combined with such an NRC-violating ``fluid". In
\cite{kn:tallandthin} and \cite{kn:singularstable} we argued, on the
basis of an analysis of possible instabilities due to pair creation
of branes [as in the work of Maldacena and Maoz \cite{kn:maoz}] that
the corrected version of spatially toral de Sitter spacetime might
resemble the metric
\begin{equation}\label{eq:G}
g\m{_c(6,\,K,\,L_{inf})_{+---} \;=\; dt^2\; -\;
K^2\;cosh^{(2/3)}\Big({{3\,t}\over{L_{inf}}}\Big)\,[d\theta_1^2
\;+\; d\theta_2^2 \;+\; d\theta_3^2]};
\end{equation}
the notation is as above and as in \cite{kn:singularstable};
L$_{\m{inf}}$ is the asymptotic inflationary length scale. This
metric is obtained by solving the Friedmann equation
\begin{equation}\label{eq:GG}
\m{\Big({{\dot{a}\over{a}}\Big)^2
\;=\;{{1}\over{L_{\m{inf}}^2}}\;-\;{{1}\over{L_{inf}^2\,a^6}}\,}},
\end{equation}
corresponding to a combination of the inflaton with a ``fluid" with
a negative ``energy density" and equation-of-state parameter 1, as
in the work of Gabadadze and Shang \cite{kn:gab}. [The special
examples of ghost-free higher curvature models discussed by Biswas
et al \cite{kn:biswas} also lead to metrics of this kind.] The
metric clearly does have a hypersurface of zero extrinsic curvature,
at t = 0, and it does very rapidly [with respect to proper time]
become indistinguishable from spatially toral de Sitter spacetime as
t increases. It can be \emph{smoothly} joined to the corresponding
Euclidean metric at t = 0, and that Euclidean metric resembles the
metric of H$^4/\bbz^3$ as soon as one moves a short distance away
from the transition zone.

To summarize, then: the Andersson-Galloway singularity theorem
dictates a violation of the NRC if the Universe is to be created
along a torus. We assume that this violation can be formally
represented by the presence of a ``fluid" which has a negative
``energy density" $\rho_{\m{NRC}}$ and some appropriate equation of
state.

\subsubsection*{{\textsf{2.4. Beyond FRW Cosmologies }}}
An obvious shortcoming of our discussion thus far is that it assumes
what we wish to prove: the local isotropy of the initial spatial
section. A theory which leads to the scale factor in equation
(\ref{eq:G}) has obviously solved the problem of divergent extrinsic
curvature, but it has not [yet] explained why the intrinsic
curvature should be so well behaved ---$\,$ the spatial sections are
spaces of constant curvature, \emph{by assumption.} Of course, a
three-dimensional space with the topology of a torus need not have
this maximally locally isotropic geometry. The generic metric on the
torus will be enormously complicated, and the characteristic
topological non-triviality of a torus will not be apparent in this
generic geometry. [Think of a very distorted two-torus: the
``handle" could be small, and be localized to some isolated region.]
It is far from clear that the wave function of the Universe will
---$\,$ or can ---$\,$ choose the maximally symmetric geometry among all
these possibilities; even recognising that the topology \emph{is}
non-trivial involves some apparently non-local mechanism.

Thus it is not true, as is sometimes said, that a ``creation from
nothing" approach \emph{automatically} solves the problem of the
Arrow of time. In fact, Hawking \cite{kn:hawking} and co-workers
\cite{kn:laf} have attempted to establish the existence of an
Arrow in the context of the original Hartle-Hawking wave
function\footnote{A much more general argument in this direction
has been given by Gibbons and Hartle \cite{kn:gibhart}; it
depends, however, on imposing a very strong positivity condition
on the Ricci tensor of the \emph{entire} spacetime, and it is
unlikely \cite{kn:maldacena} that this global condition is
actually satisfied in our Universe.}. However, it has been argued
convincingly by Kiefer and Zeh \cite{kn:zeh} [see also
\cite{kn:dab1}\cite{kn:dab2}] that this whole approach will lead
to a \emph{reversal of the Arrow of time} if the Universe should
ultimately begin to contract. This leads to all manner of apparent
internal inconsistencies, reviewed for example by Davies and
Twamley \cite{kn:twamley}. Until these objections have been fully
answered, it seems reasonable to assume that predicting a reversal
is a sign that something is seriously wrong with a given attempt
to account for the Arrow of time.

In any case, once again, \emph{all} of this work is done in the FRW
framework or small perturbations around it. It is doubtful that any
firm conclusions can be drawn from this interesting but highly
non-representative sample. To explain the Arrow of time in a
convincing way, we need to explore \emph{all possible metrics},
satisfying basic physical conditions, defined on compact
three-dimensional manifolds of given topology. [Ultimately, too, we
need to understand the consequences of choosing other initial
topologies.] That is, we need to abandon the FRW assumptions
entirely if we are really to explain the Arrow of time.

We can now formulate our ``Arrow of time strategy" concretely as
follows. We assume, in view of the preceding discussion, that the
Universe was created along a three-dimensional hypersurface with the
topology, but not necessarily the geometry, of a flat torus. The
four-dimensional sector of the Ooguri-Vafa-Verlinde wave function
describing the creation will be constructed using Euclidean
manifolds-with-boundaries. Near to the boundary, each space will
have a Euclidean metric of the form
\begin{equation}\label{eq:DODO}
g\m{(F, h)}
\;=\;\m{d\Phi^2\;+\;F^2(\Phi/L,\,\theta^c)\,h_{a\,b}\,d\theta^a\,d\theta^b}\,,
\end{equation}
where $\m{h_{a\,b}}$ is a \emph{completely general} metric on the
three-torus and where F($\Phi$/L,$\,\theta^c$) is some smooth
function such that the extrinsic curvature of the boundary vanishes.
Complexifying both $\Phi$ and L, we will obtain a Lorentzian metric
describing the earliest Universe. The two spaces are joined, as
usual, along a common three-torus. The task is then to use this OVV
wave function to show that the [overwhelmingly] most probable
spatial geometry, when a quasi-classical spacetime geometry is
selected, is an almost perfectly flat metric on this
torus\footnote{See \cite{kn:vafa} for an application of the
``entropic principle" of this general kind.}.

Unfortunately, the OVV construction is far from being sufficiently
well-understood for it to be possible to carry out this programme
directly. However, there is an indirect argument which strongly
suggests that, if it could be carried out, it would lead to the
desired answer. To understand this, let us briefly review the
general-relativistic initial value problem, in the special form it
takes for ``creation from nothing".

\addtocounter{section}{1}
\section*{\large{\textsf{3. The Initial Value Problem for ``Creation from Nothing"\footnote{To avoid confusion,
we use ($-+++$) signature in this Section.}}}} Albrecht
\cite{kn:albrecht} classifies approaches to the Arrow of time
problem into two categories: ``dynamical" and ``based on
principles". The best-known example of the latter is Penrose's
``Weyl Curvature Hypothesis" \cite{kn:penrose}; the general approach
has been ably defended by Wald \cite{kn:wald}. Such principles are
sometimes criticised on the grounds that they seem to suggest some
kind of acausal agency: how do the various parts of the initial
hypersurface ``know" that they should all have the same local
geometry?

In fact, however, such apparently acausal conditions are a
standard part of the initial value problem for field theories. For
example, Maxwell's equations do not have solutions for arbitrary
initial values of the fields and charge density: $\nabla\cdot
\bf{E}$ = 4$\pi\varrho$ and $\nabla\cdot \bf{B}$ = 0 must be
satisfied as \emph{initial value constraints}, relating the
initial electric and magnetic fields and the initial charge
density $\varrho$. To put it another way, given that the equations
have physically well-behaved solutions, these constraints are
imposed by the mathematical structure of the theory itself. This
seems promising: the fact that arbitrary initial conditions are
not permitted means that there was something ``special" about the
beginning of the Universe, and this ``special" property has
nothing to do with causal processes. It is very natural to
conjecture that the Arrow of time was established in this way:
what could be more natural than to suppose that the initial state
was fixed by the internal mathematical consistency of a
fundamental theory? The problem is that initial-value constraints
are \emph{apparently} far too weak to impose the kind of drastic
restrictions we need to obtain an Arrow. But let us examine the
question more closely.

\subsubsection*{{\textsf{3.1. Initial Value Theory for Spacetimes with Boundaries}}}

In the case of the field equations of General Relativity, the
initial value problem takes the following form \cite{kn:waldbook}.
One begins with a three-dimensional manifold $\Sigma$, on which is
given a metric h$_{\m{ab}}$, a symmetric tensor K$_{\m{ab}}$, a
function $\rho$ and a vector field J$^{\m{a}}$, together with
appropriate initial data for the ``matter fields". As in the case of
electromagnetism, the field equations will have no solutions if
these data are given arbitrarily. A solution of the field equations
with these initial data will exist ---$\,$ that is, \emph{spacetime
itself will exist} ---$\,$ only if the following equations are
satisfied:
\begin{eqnarray}\label{eq:H}
\m{D^a\,[K_{ab}\;-\;K^c_c\,h_{ab}]} & = & -\,8\pi \m{J_b}  \ \nonumber \\
              \m{R(h)\;+\;[K^a_a]^2\;-\;K_{ab}\,K^{ab}} & = &
              16\pi\rho.
\end{eqnarray}
Here D$^{\m{a}}$ is the covariant derivative operator, and R(h) is
the scalar curvature, defined by h$_{\m{ab}}$. This solution will
define a unique globally hyperbolic\footnote{A spacetime is said to
be \emph{globally hyperbolic} if it possesses a hypersurface which
intersects all inextensible timelike and null curves.} spacetime
containing matter fields [evolved forward from suitable data on
$\Sigma$] generating a stress-energy-momentum tensor T$_{\mu\nu}$.
This spacetime allows us, retrospectively, to assign interpretations
to the ``initial" data, as follows. First, $\Sigma$ is a spacelike
hypersurface with induced metric h$_{\m{ab}}$ and extrinsic
curvature K$_{\m{ab}}$. Second, we can now speak of a unit timelike
vector field n$^{\mu}$ normal to this hypersurface, and we will find
that J$^{\m{a}}$ is the projection into $\Sigma$ of the vector field
$-\,$T$^{\mu}_{\nu}\,$n$^{\nu}$. We will also find that $\rho$ is
just
\begin{eqnarray}\label{eq:I}
\rho\;=\;\m{T_{\mu\nu}\,n^{\mu}\,n^{\nu}}.
\end{eqnarray}
The vector field $-\,$T$^{\mu}_{\nu}\,$n$^{\nu}$ is the
\emph{energy-momentum flux vector} as seen by a family of
observers with unit tangent n$^{\mu}$. Therefore, $\rho$ is simply
the energy density on $\Sigma$.

Now our objective is to determine what happens if we study the
initial value problem in the case where $\Sigma$ is the boundary
[assumed to be connected, and with the topology of a
three-dimensional torus] of a smooth, paracompact
manifold-with-boundary. The following simple theorem [see
\cite{kn:milnor}, page 200] is now relevant:

\bigskip
\noindent \textsf{THEOREM [Collar Neighbourhood]: If X is a smooth
paracompact manifold-with-boundary, and $\partial$X is the boundary,
then there exists an open neighbourhood of $\partial$X in X which is
diffeomorphic to the product $\partial$X$\,\times\,$[0,1).}
\bigskip

What this means is that, near to the boundary, the space has the
product topology of a globally hyperbolic spacetime
[\cite{kn:waldbook}, pages 208-209]. Thus, the neighbourhood of the
boundary has a suitable structure for the initial-value theory. We
shall assume, in fact, that the resulting solution can be extended
to the entire manifold-with-boundary.

Now we assume, as in Section 2, that the boundary is a hypersurface
of zero extrinsic curvature. Then the constraints simplify to the
equations
\begin{eqnarray}\label{eq:J}
\m{R(h)}&=&16\pi\rho \ \nonumber\\
\m{J^a} &=&0.
\end{eqnarray}
The second equation just means that the total energy-momentum flux
vector is parallel or anti-parallel to the unit normal n$^{\mu}$.
If we adopt Gaussian normal coordinates based on $\Sigma$, then
this just states that the time-space components of the
stress-energy-momentum tensor vanish. That is manifest for the
inflaton tensor given in (\ref{eq:A}), since those components
reverse sign with time, and $\Sigma$ is a hypersurface of time
symmetry. A similar argument applies to the other contributions to
the time-space components.

There is a more interesting way of interpreting the second equation.
Take any smooth manifold-with-boundary X, and let m be a point on
the boundary, $\partial$X. The tangent space to X [\emph{not}
$\partial$X] at m is naturally divided into two subsets, defined as
follows [see \cite{kn:milnor}, page 200]. Let V be a tangent vector
to X, but not to $\partial$X, at m. Then V is said to \emph{point
inwards} if V is the tangent vector at m of a smooth curve
$\gamma\,:\,[0,\,\epsilon)\,\rightarrow\,$ X with $\gamma$(0) = m.
Vectors pointing \emph{outwards} are defined analogously, using
curves of the form $\gamma\,:\,(-\,\epsilon,\,0]\,\rightarrow$ X,
again with $\gamma(0)$ = m.

Let us choose the unit normal n$^{\mu}$ to point inwards. Then the
second equation in (\ref{eq:J}) just says that the energy-momentum
flux vector on $\Sigma$ either points perpendicularly into
spacetime or out of it [or is zero]. But clearly, if the classical
or quasi-classical Universe is being \emph{created} from
``nothing", then the energy-momentum flux vector on $\Sigma$
cannot point ``outwards" from spacetime. We therefore conclude
that, in terms of classical language,
\begin{eqnarray}\label{eq:K}
\rho\;=\;\m{\rho_{inf}\;+\;\rho_{NRC}\;\geq\;0},
\end{eqnarray}
everywhere on $\Sigma$; here $\rho_{\m{inf}}$ is the [positive]
energy density of the inflaton, and $\rho_{\m{NRC}}$ is the
[negative] effective energy density discussed in the previous
section.

This is obviously a reasonable condition in the quasi-classical
domain, but we ask the reader \emph{not} to think of it in terms
of the familiar energy conditions [such as the dominant or weak
energy conditions] ---$\,$ though of course it does have the
interpretation that ``normal" matter, represented by
$\rho_{\m{inf}}$, cannot be outweighed by the exotic NRC-violating
``fluid" represented formally by $\rho_{\m{NRC}}$. Rather, it
expresses the idea that, whatever it may do in the interior of
spacetime, the energy-momentum flux vector of a quasi-classical
spacetime should not, when evaluated on the boundary, point in a
``direction" which literally does not exist. In particular, we do
not insist that (\ref{eq:K}) should hold throughout spacetime,
away from the boundary. We also do not require it to be true of
\emph{all} of the spaces surveyed by the OVV wave function.
Instead, it is to be imposed only on the ``initial" boundary of
that spacetime which is selected as constituting the
quasi-classical world. [The precise way in which this selection is
done is of course a deep question which we shall not consider
here: see for example \cite{kn:gell} and the discussion of the
``emergence of classicality" in Section 7 of \cite{kn:albrecht}.]

To summarize, then: from the point of view of the initial value
problem, the only special condition imposed by ``creation from
nothing" is that the [total] quasi-classical energy-momentum flux
vector should not be allowed to point outwards from the boundary.
That imposes (\ref{eq:K}), which is, or appears to be, a very mild
restriction indeed. This is to be fed into the first member of
(\ref{eq:J}), as the only restriction on the geometry of $\Sigma$.

\subsubsection*{{\textsf{3.2. Can The Constraints Always Be Satisfied?}}}

In the case of the initial value problem for electromagnetism, the
initial value constraints do not merely restrict the relationships
between the initial fields and their sources: they also forbid
certain kinds of situations altogether. To take a relevant example:
on a flat torus, the integral version of Gauss' law shows
immediately that the electric charge density cannot be everywhere
strictly positive, and the only way it can be everywhere
non-negative is by being identically zero. This is still, however, a
very weak condition on the electric field, since there are clearly
infinitely many distinct vector fields having $\nabla\cdot \bf{E}$ =
0. The problem, of course, is that the divergence is a \emph{sum} of
terms which need not, individually, have any particular sign, so the
total can easily be zero.

In view of this, it would appear that our condition (\ref{eq:K}),
which just requires that the metric on the initial [boundary]
hypersurface $\Sigma$ should satisfy
\begin{eqnarray}\label{eq:L}
\m{R(h)\;\geq\;0},
\end{eqnarray}
does not restrict the geometry of $\Sigma$ very strongly. For the
scalar curvature is essentially just the \emph{sum} of the curvature
tensor components, and for a distorted metric on the sphere or the
torus one has no grounds for predicting a preponderance of one sign
over the other. [Thus for example one can construct metrics on S$^3$
with enough negative curvature components so that the \emph{total}
is negative, although of course some components are positive; see
\cite{kn:porrati}]. In fact, since there is no analogue of Gauss'
theorem here, we expect the restriction to be even milder than in
the case of electromagnetism. Nevertheless, the relation between the
scalar curvature and the metric is very intricate, so we should not
jump to the conclusion that (\ref{eq:L}) imposes no significant
restriction on the metric.

Let us consider the case where $\Sigma$ has the topology of a
three-dimensional sphere, as in the original Hartle-Hawking
construction \cite{kn:hartle}. The question now is this: given a
function S(x) on S$^3$, which is nowhere negative but otherwise
arbitrary, can one find a metric h$_{\m{ab}}$ on S$^3$ such that the
scalar curvature of h$_{\m{ab}}$ is precisely equal to S(x)?

Intuitively, one might think that the answer should always be
positive. We have, after all, all six independent components of
h$_{\m{ab}}$ at our disposal, and only one condition to be
satisfied. If this is so, then of course it means that (\ref{eq:K})
alone is an exceedingly weak constraint. It would also mean that the
round metric on S$^3$, and small perturbations of it, constitute an
infinitesimal minority of the possible metrics satisfying
(\ref{eq:L}); one would then conclude that perturbations around the
corresponding FRW spacetime probably do not adequately explore the
set of all spacetimes with topologically spherical initial data.

Surprisingly, the answer to this question is known.

\subsubsection*{{\textsf{3.3. The Kazdan-Warner Classification}}}
The question we have raised here is a particular case of the
following problem. Let M be a smooth compact manifold [without
boundary] and let S(x) be a given smooth function on M. Let
h$_{\m{ab}}$ be a metric on M with scalar curvature R(h). Can the
second-order partial differential equation
\begin{eqnarray}\label{eq:M}
\m{R(h)\;=\;S(x)},
\end{eqnarray}
be solved for h$_{\m{ab}}$?

The answer to this question takes a rather strange form. Instead of
depending mainly on the details of S(x), it depends primarily on the
[differential] topology of M, and takes the form of a classification
of smooth compact manifolds [without a fixed metric]. This very
remarkable classification is given by the \emph{Kazdan-Warner
Theorem}, as follows \cite{kn:kazdan}:

\bigskip
\noindent \textsf{THEOREM [Kazdan-Warner]: All compact manifolds of
dimension at least three fall into precisely one of the following
three classes:}

\medskip

\noindent \textsf{[P] On these manifolds, every smooth function is
the scalar curvature of some Riemannian metric.}

\medskip
\noindent \textsf{[Z] On these manifolds, a smooth function can be a
scalar curvature of some Riemannian metric if and only if it either
takes a negative value somewhere, or is identically zero.}

\medskip
\noindent \textsf{[N]  On these manifolds, a smooth function can be
a scalar curvature of some Riemannian metric if and only if it takes
a negative value somewhere.}

\bigskip

For example, spheres obviously do have a metric of strictly positive
scalar curvature; since the theorem classifies all compact manifolds
of dimension at least three, it follows that all spheres of
dimension at least three lie in KW class P; this means that equation
(\ref{eq:M}) always has a solution for \emph{any} S(x), no matter
how intricate it may be, on S$^3$. \emph{Any} initial distribution
of energy, positive, negative, or mixed, is compatible with the
initial-value constraints: just choose the three-dimensional metric
to be the corresponding solution of the scalar curvature equation.

As we feared, requiring that the energy density should be
non-negative along the initial hypersurface $\Sigma$ is an even
weaker condition, \emph{when the initial spatial topology is
spherical}, than requiring non-negative electric charge density. But
the situation is very different in the case of toral topology.

\addtocounter{section}{1}
\section*{\large{\textsf{4. The Amazing Torus}}}
It is clear that the KW class of T$^3$ cannot be N. The question is
whether it is P or Z. We remind the reader that intuitions based on
the canonical metrics [or on the two-dimensional case, where the
Gauss-Bonnet theorem applies] are highly misleading: there certainly
are metrics of strictly negative scalar curvature on both S$^3$ and
T$^3$ ---$\,$ indeed, the Kazdan-Warner theorem implies that there
are such metrics on \emph{all} compact three-dimensional manifolds.

In fact, the question of the KW class of the torus was settled only
relatively recently, by extremely deep results due to Schoen, Yau
\cite{kn:schoenyau}, Gromov, and Lawson \cite{kn:lawson}. Let us
examine the global geometry of T$^3$ more closely.

We can define T$^3$ abstractly as the quotient $\bbr^3/\bbz^3$,
where one can think of $\bbz^3$ as a \emph{lattice}, that is, the
additive group of vectors with integer components relative to some
fixed basis; if for example we take this basis to be the canonical
orthonormal basis in $\bbr^3$, then we are dealing with a cubic
torus. Now let n be \emph{any} positive integer, and define
(n$\,\cdot\,\bbz$)$^3$ in the obvious way. Then
$\bbr^3$/(n$\,\cdot\,\bbz$)$^3$ is clearly a covering manifold of
the original torus; it is an n$^3$-fold covering. Thus we see that
we can find \emph{arbitrarily large} coverings of T$^3$, since n can
be as large as we please. [By contrast, the real projective space
$\bbr$P$^3$ can be ``made larger" by taking a covering, but not
\emph{arbitrarily} larger.] This notion can be generalized [by
defining ``arbitrarily large" in terms of the amount by which the
lengths of tangent vectors are contracted by the covering map]. One
says that manifolds like T$^3$ having this property are
\emph{enlargeable}. [See \cite{kn:lawson}, page 302, for the formal
definition.]

``Enlargeability" has several interesting properties: for example,
the connected sum\footnote{The connected sum of two manifolds of the
same dimension is formed by deleting the interior of a ball in each,
and joining the resulting spaces with a cylinder.} of any manifold
with an enlargeable manifold is again enlargeable. Many compact
three-manifolds are enlargeable: in particular, any manifold
admitting a metric of non-positive sectional curvature is
enlargeable; thus, tori [and their non-singular quotients] are very
special representatives of this large class.

The work of Schoen, Yau \cite{kn:schoenyau}, Gromov, and Lawson
[\cite{kn:lawson}, page 306] can be summarized as follows:

\bigskip
\noindent \textsf{THEOREM [Schoen-Yau-Gromov-Lawson]: There is no
metric of positive scalar curvature on any compact enlargeable spin
manifold.}
\bigskip

The proof of this theorem is deep, as is evidenced by the appearance
of the ``spin" condition\footnote{Recall that an orientable
Riemannian manifold is said to be \emph{spin} if the SO(n) structure
group of its bundle of orthonormal frames can be lifted to
Spin(n).}; while apparently irrelevant, it is in fact necessary. [In
three dimensions, however, it is automatically satisfied.]

It follows from this theorem that compact enlargeable spin manifolds
can never be in Kazdan-Warner class P. Now tori are compact,
enlargeable, and spin; hence, we conclude that all tori are in KW
class Z.

We see now, from the Kazdan-Warner theorem, that the only way for
condition (\ref{eq:K}) to be satisfied is by having
\begin{eqnarray}\label{eq:N}
\rho\;=\;\m{\rho_{inf}\;+\;\rho_{NRC}\;=\;0},
\end{eqnarray}
everywhere on the initial torus. The situation is therefore
analogous to the case, discussed earlier, of the Maxwell initial
value constraint $\nabla\cdot \bf{E}$ = 4$\pi\varrho$: we saw that,
on a torus, $\varrho\;\geq\;0$ actually implies $\varrho\;=\;0$. As
in that case, however, this appears to be a weak constraint,
analogous to $\nabla\cdot \bf{E}$ = 0, on the metric: as we have
repeatedly emphasised, the scalar curvature is nothing but the
\emph{total} of all of the curvature components, so we expect that
it should be easy to find \emph{many} metrics of vanishing scalar
curvature on the initial torus.

This, however, is where the analogy with the electromagnetic case
breaks down; and it breaks down spectacularly. For Gromov and
Lawson, extending a theorem of Bourguignon, were able to prove
[\cite{kn:lawson}, page 308] the following theorem.

\bigskip
\noindent \textsf{THEOREM [Bourguignon-Gromov-Lawson]: If a metric
on a compact enlargeable spin manifold has zero scalar curvature,
then that metric must be exactly locally flat, that is, the
curvature tensor must vanish identically everywhere on the
manifold.}
\bigskip

\noindent This is an astonishing result: the vanishing of a
\emph{sum} of curvature components, the scalar curvature, forces
\emph{each} curvature component to vanish separately on these
manifolds.

We now see that, in sharp contrast to the case of ``creation from
nothing" on S$^3$, the initial value constraints in the case of
topologically toral initial data are enormously powerful: \emph{the
only way in which (\ref{eq:K}) can possibly be satisfied is if the
initial torus is, classically, perfectly flat.}

We claim that this is the origin of the initial ``specialness" from
which the Arrow of time derives. The internal mathematical
consistency of the OVV theory, in the form of the initial value
constraints, forces the Universe to be created in an extremely
non-generic initial state: the initial spatial section has to be
locally isotropic, with the consequences explained in Section 1.

More generally, we can ask: what happens if we assume that the
Universe was created along a compact three-dimensional space with
some non-toral topology? This is a question which could not have
been fully answered until very recently; but by combining one of the
recent celebrated results due to Perelman \cite{kn:perelman} with
the remarkable work of Gromov and Lawson, we can now do so: for
these theorems allow us to specify precisely which compact
three-dimensional manifolds belong to the respective KW classes.
Briefly, the classification runs as follows. [We confine ourselves
to the case of orientable spaces; the extension to the
non-orientable case is interesting, but raises no issues relevant to
our discussion here.]

Milnor's \cite{kn:milnor2} ``prime decomposition" theorem states
that any compact orientable 3-manifold M$^3$ can be expressed as a
connected sum in the following way:
\begin{equation} \label{eq:O}
\m{M^3 = P_1 \,\#\, P_2 \, \#\, ... \,\#\, (S^1 \times S^2)\, \# \,
(S^1 \times S^2)\, \# \,... \,\#\, K_1 \,\# \,K_2 \,\#\, ...,}
\end{equation}
where each P$_{\m{n}}$ is a compact manifold with a finite
fundamental group, where $\#$ denotes the connected sum, and where
each K$_{\m{n}}$ is an Eilenberg-MacLane space of the
form\footnote{A K$(\pi,1)$ space is just a compact manifold whose
only non-trivial homotopy group is its fundamental group.}
K$(\pi,1)$. Now it is known [\cite{kn:lawson}, page 324] that no
K$(\pi,1)$ can accept a metric of positive scalar curvature, and
that the same is true of the connected sum of a K$(\pi,1)$ with any
other compact manifold. [This gives an alternative explanation of
the fact that the torus has KW class Z.] Therefore, if M$^3$ has KW
class P, there can be no K$_{\m{n}}$ factors in its Milnor
decomposition. Furthermore, Perelman's proof \cite{kn:perelman} of
the \emph{elliptization conjecture} means that each P$_{\m{n}}$ is
diffeomorphic to a manifold of the form S$^3/\Gamma$, where $\Gamma$
is a finite group drawn from a completely known list [given, for
example, in \cite{kn:wolf}\cite{kn:weeks}]. Next, it is known
\cite{kn:schoen} that the connected sum of two compact manifolds
admitting metrics of positive scalar curvature likewise admits a
metric of positive scalar curvature. Finally, it can be shown
[\cite{kn:lawson}, page 308] that if a manifold is in KW class Z,
then any metric of zero scalar curvature on that manifold must in
fact be Ricci-flat; in three dimensions, that means that it must be
flat.

Combining all these results, we have the following classification.
Let M$^3$ be a compact three-dimensional orientable manifold. Then
the following statements hold.

\bigskip
$\bullet$ $\;\;$ M$^3$ has KW class P if and only if it can be
expressed as
\begin{equation} \label{eq:P}
\m{M^3 = S^3/\Gamma_1 \,\#\, S^3/\Gamma_2 \, \#\, ... \,\#\, (S^1
\times S^2)\, \# \, (S^1 \times S^2)\, \# \, ...,}
\end{equation}
where each $\Gamma_{\m{i}}$ belongs to an infinite but completely
known list of finite groups.

\bigskip
$\bullet$ $\;\;$ M$^3$ has KW class Z if and only if it has the
topology of one of the compact orientable \emph{platycosms},
listed in \cite{kn:wolf}\cite{kn:conway}; these are manifolds of
the form T$^3$/$\Omega$, where $\Omega$ is one of six possible
finite groups.

\bigskip
$\bullet$ $\;\;$ All other compact orientable 3-dimensional
manifolds are in KW class N.

\bigskip

We see that the members of KW class P are very special, while class
Z is so special that its members can be given explicitly in a finite
[and short] list. In this sense, the generic compact
three-dimensional manifold is in KW class N. Note that the fact that
compact spaces of constant negative curvature are enlargeable,
together with known facts about the topology of such spaces, imply
that they are in class N; see page 306 of \cite{kn:lawson}.

Now it is easy to see that everything we said earlier regarding
S$^3$ applies equally to \emph{every} manifold in KW class P. For
all of these spaces, there are spacetimes of arbitrary complexity
which can, via the initial value constraints, satisfy condition
(\ref{eq:K}), and so we cannot expect the initial value constraints
to yield an Arrow of time if the Universe is created along a space
with such a topology.

At the other extreme, we have the enormous collection of
three-dimensional manifolds in KW class N; but \emph{none of these}
has \emph{any} metric of non-negative scalar curvature. According to
the initial value constraint equations, then, they can never satisfy
condition (\ref{eq:K}), under any circumstances. One can for example
take a compact manifold of constant negative curvature, and deform
it in an arbitrarily complicated way, but it will never be able to
satisfy (\ref{eq:K}). The apparently innocuous condition
(\ref{eq:K}) has turned out to be very powerful: it rules out all of
the many three-dimensional manifolds with topologies putting them in
KW class N.

The only survivors are the six [orientable] members of KW class Z.
These are, in Conway's [prescient] ``cosmic" terminology
\cite{kn:conway}, the torus or torocosm T\3, the dicosm
T\3/$\bbz_2$, the tricosm T\3/$\bbz_3$, the tetracosm T\3/$\bbz_4$,
the hexacosm T\3/$\bbz_6$, and the didicosm or Hantzsche-Wendt space
T$^3$/[$\bbz_2\;\times\;\bbz_2$]. These three-dimensional compact
orientable spaces, and these alone, can lead to an Arrow of time in
the way we have discussed here.

\addtocounter{section}{1}
\section*{\large{\textsf{5. Some Consequences}}}
The solution of the Arrow of Time problem advocated here differs
very greatly from previous proposals, and this in several ways. Some
of these are obvious: the non-reliance on rare fluctuations, the
reliance on deep topological-geometric properties of the torus and
its descendants, and so on. Others are less obvious, and will be
briefly discussed here.

\subsubsection*{{\textsf{5.1. Remarks on Time Orientation}}}
We have asserted that the crucial condition (\ref{eq:K}) [which in
fact reduces to equation (\ref{eq:N})] does \emph{not} represent an
energy condition, and in fact we only insist that it should hold on
the initial hypersurface. Actually, however, the term
$\rho_{\m{NRC}}$ in equation (\ref{eq:N}) decays away very rapidly
with the expansion: it was argued in \cite{kn:singularstable} that
this rapid decay is demanded as a by-product of conditions ensuring
``stringy" non-perturbative stability, such as those discussed in
\cite{kn:porrati}. In fact, the decay is probably as rapid as
causality allows; if we assume an approximately constant
equation-of-state parameter, the decay is according to the inverse
sixth power of the scale factor. The specific metric obtained in
that case is in fact precisely $g\m{_c(6,\,K,\,L_{inf})_{+---}}$,
given in equation (\ref{eq:G}), as one can see from equation
(\ref{eq:GG}). This metric gives an approximate description of the
spacetime geometry during the pre-inflationary era; we remind the
reader that it rapidly approaches the metric of Spatially Toral de
Sitter spacetime, that is, the inflationary metric.

It should be noted that the total energy density in this spacetime
is given by the simple expression
\begin{equation}\label{eq:R}
\m{\rho \;=\;{{3}\over{8\pi
L_{\m{inf}}^2}}\,tanh^2\Big({{3\,t}\over{L_{inf}}}\Big)}\;\geq\;0.
\end{equation}
The inequality will remain valid even if we do not model the
inflaton by a cosmological constant, since the inflaton energy
density will certainly decay much more slowly than the negative
term. Thus, while (\ref{eq:K}) is not itself an energy condition,
it does \emph{imply} a [very weak] condition of this sort: the
total energy density as seen by the fundamental observers is
always non-negative. That is, for these observers, the total
energy-momentum flux vector points, for all t $>$ 0, towards [what
we may now justly call] the future. This vector field establishes
a time-orientation for the early Universe.

In short, the theory advanced here establishes a well-behaved flow
of time throughout the earliest era of the Universe, not just at t =
0 itself. The importance of establishing this time orientation in
any theory of the Arrow of time is discussed in detail by Aiello et
al \cite{kn:aiello}.

\subsubsection*{{\textsf{5.2. The Shape of the Earliest Universe}}}
By using the methods of global differential geometry explained in
the preceding section, we have shown that the only way to satisfy
condition (\ref{eq:K}) is for the Universe to be ``created from
nothing" along a boundary which is perfectly isotropic at every
point. The standard arguments of classical cosmology now imply that
the spacetime metric is, initially, an exact FRW metric. \emph{Thus,
the FRW family of metrics with flat [but compact] spatial sections
are good models of spacetime at very early, as well as at very late
times.} In particular, the spacetime with the metric
$g\m{_c(6,\,K,\,L_{inf})_{+---}}$ we discussed earlier may be a more
accurate representation of the earliest spacetime geometry than we
originally supposed. The details of this geometry are therefore of
some interest; let us mention a few unusual features.

The most important question in this regard is the value of the
parameter K, which fixes the size of the Universe at its birth. In a
string theory formulated on a torus, T-duality implies that by far
the most natural initial length scale would be the string length
scale. This is normally taken to be considerably larger than the
Planck scale. Let us assume that K is indeed given by the string
scale.

This scale is independent of the inflationary length scale,
L$_{\m{inf}}$, which is roughly two orders of magnitude larger
than K. By computing the full extent of conformal time
\cite{kn:tallandthin}, which we denote by $\Omega$, one can show
that the Penrose diagram [Figure 1] will be a rectangle which is
roughly 100 times as high as it is wide. [The width of the diagram
is $\pi$, corresponding to any one of the angular coordinates on
the torus.] The horizontal line at the bottom of Figure 1
represents the creation of this universe at proper time t = 0,
while the upper horizontal line is future conformal infinity, that
is, physically, the end of the inflationary era.

\begin{figure}[!h]
\centering
\includegraphics[width=0.3\textwidth]{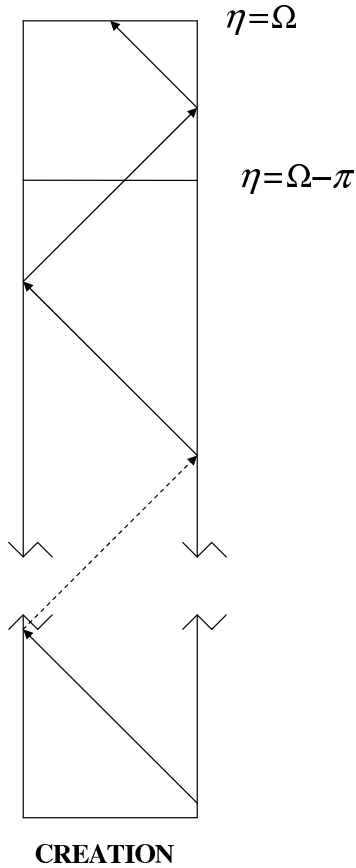}
\caption{Penrose diagram corresponding to
$g\m{_c(6,\,0.01\,L_{inf},\,L_{inf})_{+---}, \,t \geq}$ 0.}
\end{figure}

The principal consequence of this is that the early Universe was
\emph{small}, in the technical sense that circumnavigations were
easily performed. That is, all parts of the spatial sections are in
causal contact [symbolized in the diagram by the worldline of a
photon circumnavigating the torus] until nearly the top of the
diagram. This means, as Linde points out \cite{kn:lindetypical},
that \emph{chaotic mixing} \cite{kn:mixing} will prevent the growth
of perturbations until such circumnavigations cease to be possible.
This happens at about $\pi$ units of conformal time below the top of
the the diagram; at that time, the spatial sections will, because of
chaotic mixing, still be extremely regular. Thus, the FRW structure
continues to be an excellent approximation throughout this period.
One might prefer to regard the point which is $\pi$ units below the
top of the diagram as the ``start" of Inflation, since by that time
the Universe has expanded to a size comparable to L$_{\m{inf}}$; so
that the earlier period is the ``pre-inflationary" era during which
the conditions for Inflation to begin, in the conventional sense,
were prepared.

In summary, the theory of the Arrow presented here leads to a
definite picture of the geometry of the earliest Universe. It is a
FRW geometry with an unusual causal structure. This causal structure
is such that the non-trivial topology of the spatial sections would
have been very much in evidence in the pre-inflationary era. If the
sections have non-trivial holonomy groups\footnote{These are the
groups generated by parallel transport around closed loops. Even for
flat manifolds, parallel transport around non-contractible loops can
be non-trivial.}, for example, this could have physical
consequences. This observation may ultimately allow us to
distinguish physically between the various ``platycosms" in KW class
Z [all of which, apart from the torus, have finite non-trivial
holonomy groups].

\subsubsection*{{\textsf{5.3. The Weyl Curvature Hypothesis}}}
We argued above that a FRW metric like
$g\m{_c(6,\,K,\,L_{inf})_{+---}}$, given in (\ref{eq:G}), yields a
good approximation of the earliest spacetime geometry. Now
$g\m{_c(6,\,K,\,L_{inf})_{+---}}$ can be expressed in the following
form:
\begin{equation}\label{eq:Q}
g\m{_c(6,\,K,\,L_{inf})_{+---} \;=\; L_{inf}^2\,G(\eta_+)^2\,[\,
d\eta_+^2\;-\;d\theta_1^2 \;-\; d\theta_2^2 \;-\; d\theta_3^2]},
\end{equation}
where $\eta_+$ is conformal time, defined on a certain finite
interval [of length fixed by L$_{\m{inf}}$/K], and where G($\eta_+$)
is a certain function which diverges at a finite value of $\eta_+$,
which represents the infinite future in proper time. The metric is
manifestly conformally flat.

Penrose \cite{kn:penrose} has suggested that the Arrow of time is
due to the vanishing of the Weyl tensor at initial, and only at
initial, cosmological singularities. The present work may be
regarded as an attempt to provide a rationale for the Weyl
Curvature Hypothesis [adapted to the case of ``creation from
nothing"] by embedding it in string theory. We should note,
however, that our version is much stronger than Penrose's. For
since all FRW metrics have zero Weyl curvature, the latter would
allow \emph{any} FRW metric near the creation. Here we have been
led to FRW models with necessarily flat spatial sections, with
metrics which are \emph{globally} conformal to [spatially
compactified] Minkowski spacetime.

It may be of interest at this point to recall that the
Euclidean-to-Lorentzian transition is performed here by
complexifying \emph{conformal} time: so, up to a global conformal
factor, the complexification here is just the usual Wick rotation of
flat-spacetime quantum field theory. Perhaps this will answer some
or all of the objections to performing complexifications of time
coordinates on curved spacetimes.

To summarize: we argue that Penrose's ``Weyl Curvature Hypothesis"
can be naturally embedded in, and justified by, the OVV version of
``creation from nothing". However, so far we have only explained why
the Weyl curvature should be expected to vanish at the
\emph{beginning} of time. Penrose claims that we should \emph{not}
expect it to vanish at the end [if any]. We now explain how this
claim, too, can be justified by our approach.

\subsubsection*{{\textsf{5.4. Pleading Innocent to Cosmic Hypocrisy}}}
Price \cite{kn:price} has argued very persuasively that most
accounts of the Arrow of time are guilty of a ``double standard":
they make use of arguments about the beginning of time that seem to
apply equally to its \emph{end}. This leads to conclusions that few
can accept. A well-known example is Inflation: if it is ``natural"
[in the sense of ``requiring no specific mechanism"] for it to occur
at the beginning, then it must be equally natural for ``deflation"
to occur should the Universe eventually contract. Most authors find
this hard to believe.

Here we have argued that Inflation is not ``natural" in this sense:
it only occurs because of the demands of the internal mathematical
consistency of string theory, as revealed by the
Bourguignon-Gromov-Lawson theorem. Whether ``deflation" will occur
if the Universe eventually contracts is a matter to be settled in
the same way.

Before we begin, let us note that it is far from certain that string
theory even \emph{permits} eventual contraction. If the current
cosmic acceleration is due to a true cosmological constant, there
will be no contraction; and some have argued \cite{kn:polchinski}
that string theory favours a cosmological constant as the
explanation of the acceleration. However, let us leave this to one
side, and consider universes which do re-contract, since this will
throw some light on basic questions regarding the Arrow. In
particular, let us consider cosmological models which, like those
discussed by Maldacena and Maoz \cite{kn:maoz}, actually have a
small \emph{negative} cosmological constant.

A negative cosmological constant, however small in magnitude, will
eventually halt the expansion; for all other fields, including
[especially] the NRC-violating ``fluid", will be insignificant when
the Universe is extremely large. The subsequent contraction,
however, will eventually revive those fields, and the repulsion
generated by the ``fluid" will finally dominate and halt the
contraction. That is, a final singularity will be averted in the
same way as the initial singularity. The result will be a ``bounce",
as in the work of Kachru and McAllister \cite{kn:kachru} [see also
\cite{kn:stach}], along a spacelike hypersurface of zero extrinsic
curvature.

We now claim that the irregularities which developed during the
expansion will grow rapidly as the Universe contracts; so the
geometry at the bounce, while not singular, should be extremely
irregular. That is, entropy [of all kinds] will continue to increase
up to, and beyond, the bounce.

The question Price would raise at this point is this: can this claim
be internally consistent? Why does the mechanism which enforced low
entropy at the creation not enforce it at the bounce?

The answer has two parts. First, the initial boundary is very
different from the bounce hypersurface: \emph{the latter has both
a past and a future}. Therefore we have no motive, in this case,
for enforcing the requirement that the total energy-momentum flux
vector should point in any particular direction, when evaluated on
that hypersurface. Since the bounce itself is due to the dominance
of the negative ``energy density" $\rho_{\m{NRC}}$, there is no
contradiction in assuming that the total flux vector points in one
direction at some points on the hypersurface, and in the opposite
direction at others: that is, the total density need not be
non-negative everywhere on the hypersurface.

However, this argument seems to suggest that the asymmetry here is
based on a claim that the ``end" of the Universe is objectively
different from its beginning. There \emph{is} such a difference
---$\,$ the ``initial" hypersurface does not have a past ---$\,$ but
this is not the real crux of the matter. To understand what is
happening here, let us proceed as follows. We saw that the
contraction is ultimately halted, along a hypersurface of zero
extrinsic curvature, by the dominance of the negative quantity
$\rho_{\m{NRC}}$. Let us assume, in order to maintain an exact
logical symmetry with the situation at the beginning of time, that
the total density is non-positive everywhere at the bounce.

This brings us to the real point: we know that non-negative energy
density on a topologically toral hypersurface of zero extrinsic
curvature has drastic consequences for the intrinsic geometry. What
are the consequences of non-positive energy density? A contradiction
will indeed arise if, as one might expect on symmetry grounds,
non-positive energy density has similar consequences to non-negative
energy density.

Since we are dealing, once again, with a hypersurface of zero
extrinsic curvature, the initial value constraints again reduce this
to a question about scalar curvature. The answer to our question is
then given by the Kazdan-Warner theorem: since the torus and its
quotients are in KW class Z, \emph{every} smooth non-positive
function, no matter how convoluted, can be the scalar curvature of a
[similarly convoluted] metric on these spaces. [In fact, the same
conclusion holds even if there are \emph{some} points where the
function is positive.] It follows that there is no contradiction in
assuming that the spatial sections have increasingly distorted
geometry as the bounce is approached.

Ultimately, then, the reason that the Universe can have a
low-entropy beginning and a high-entropy ``end" [the bounce] can
be traced back to \emph{the asymmetry in the set of all possible
metrics on the torus}. The torus ---$\,$ in sharp contrast to the
sphere, on which ``anything goes" ---$\,$ favours one sign of
scalar curvature over the other. This is how it is possible for
one ``end" of the Universe to be more irregular than the other. It
is very striking that this extremely deep property of the set of
all possible toral geometries accounts for such a basic property
of our Universe.

The case of the interior of black holes is also instructive. If
the Arrow of time is somehow related to the increase or decrease
of the \emph{volume} of regions of space, then one is entitled to
ask, as Price does \cite{kn:price}, whether the Arrow of time
reverses inside the black hole event horizon. The answer here is
clear: the Arrow does \emph{not} reverse. Consider a black hole,
formed from the collapse of a star, subject to Cosmic Censorship.
That is, the singular region [or whatever replaces it in string
theory] is spacelike, and the black hole does not affect the
topology of the spatial sections. Now one hopes that
string-theoretic effects somehow resolve the singularity, and this
may well impose special conditions on the parts of spatial
sections which are inside the event horizon. Whatever these
conditions may be, however, they do not control the geometry of
those parts of a spatial section which lie far outside the
horizon; hence there is, here, simply no analogue of the
\emph{global} vanishing of the extrinsic curvature, or of the
condition (\ref{eq:K}) which we imposed \emph{globally} on the
initial hypersurface. Hence there is no reason to think that the
Arrow of time anywhere, including the interior of the event
horizon, is affected by the unusual geometry near to [whatever
replaces] the classical singularity.

Price's \cite{kn:price} discussion of these questions is valuable,
because it makes it very clear that no \emph{local} mechanism can
give an account of an Arrow of time which never reverses. As Price
insists, whatever caused the initial state of the Universe to be
so regular might well act in the same way on a black hole
singularity; and his argument has the more force in this case, in
that the absence of any evidence for this possibility is, for
obvious reasons, not surprising. The claim is that conditions
\emph{near to} the singularity of a [Cosmically Censored] black
hole are similar to those \emph{near to} a generic cosmological
singularity. This is reasonable, but it does not allow us to
conclude that the Arrow reverses inside a black hole, because the
Arrow is, we claim, a truly \emph{global} property of the
Universe. It is only when we consider the geometry of an entire
compact space that we find strange phenomena such as the extreme
\emph{asymmetry} of the space of all metrics on the three-torus.

In summary: in the present theory, the Arrow of time is correlated
with the cosmic expansion only in the sense that both have a
common origin, the condition (\ref{eq:K}) which just expresses the
idea that the energy-momentum flux vector should point inwards at
the boundary of spacetime ---$\,$ that is, the idea that the
Universe was created from \emph{nothing}. [As we explained above,
this condition implies (\ref{eq:N}). While the total energy
density vanishes initially, the total pressure does not; with a
causal equation of state for the NRC-violating ``fluid", the
initial total pressure must have been negative. The Raychaudhuri
equation shows that this ``primordial pressure" forces the
Universe to begin expanding as soon as it is born.] In the
unlikely event that the Universe should begin to contract, that
correlation will be broken: the Arrow of time never reverses.

\addtocounter{section}{1}
\section* {\large{\textsf{6. Conclusion}}}
The existence of an Arrow of time is the most wildly improbable
feature of our Universe. It is therefore the feature most urgently
in need of explanation. We have proposed an explanation here, one
based ultimately on the extraordinarily atypical and asymmetrical
structure of the space of Riemannian metrics on the torus. These
special properties of the torus are communicated to the inflaton,
putting it into its lowest-entropy state, by constraints imposed by
the internal mathematical consistency of the theory. Much remains to
be done to complete this explanation: in particular, of course, we
need a much better understanding of the OVV wave function. Let us
conclude by mentioning some less apparent issues.

First, an essential step in our argument exploits the detailed
structure of the initial value problem for gravity. Note that this
is still by no means a closed subject; work continues on
understanding it, particularly in the most physically interesting
case [scalar fields interacting with gravity]; see for example
\cite{kn:bruhat}. Further study of this problem may well be of great
interest in connection with the Arrow.

Secondly, our main technical resource in this work has been the
global differential geometry of the scalar curvature invariant. This
is a well-developed subject, which, however, is also an area of
active research; see \cite{kn:topologically} for mathematical
references and a possible physical application. Understanding the
ways in which topology constrains geometry is the problem at the
core of modern global differential geometry, and it is likely to
have further physical implications.

Finally, the following speculation is suggested by this work.

It has often been claimed \cite{kn:susskind} that ``baby universes"
can branch off from a given Universe. \emph{If} these babies can be
sufficiently similar to our observed Universe, then we could be
living in one of the babies.

It is one thing to produce a baby; quite another, however, to ensure
that the baby will behave in the way that one wishes. In this work,
we have suggested that the Arrow of time is a consequence of the
extremely special conditions that arise when a [toral] Universe is
created from \emph{nothing}. Baby universes, by definition, are
\emph{not} created from nothing; they arise in quite a different
way. \emph{We therefore conjecture that there is no Arrow of time in
any baby universe.}

Evidence for this conjecture will be presented elsewhere.
Meanwhile, we leave the reader with the following observation.
While the conditions for the existence of sentient life are
notoriously controversial, surely all can agree that it does not
exist in conditions of thermal equilibrium\footnote{Here we are
neglecting the possible existence of ``Boltzmann brains". See for
example \cite{kn:page}\cite{kn:carlip}\cite{kn:sred} for recent
interesting discussions of this question.}. If, as could easily be
the case, it proves to be extremely difficult to establish an
Arrow of time in a baby universe, then the Landscape
\cite{kn:bousso} might be very drastically depopulated. If, for
example, baby universes \emph{never} have an Arrow, then we, the
inhabitants of \emph{this} Universe, may perhaps be alone in the
``multiverse". Whether or not this is so, the following general
point should be clear: \emph{nothing can be said as to whether a
``multiverse" can be populated until the Arrow of time is
thoroughly understood.} It is to be hoped that this realization
will spur further interest in this perplexing problem.

\addtocounter{section}{1}
\section*{\large{\textsf{Acknowledgement}}}
The author is grateful, as ever, to Wanmei for the diagram, and for
other gifts too numerous to be listed here.


\begin{thebibliography}{18}

\bibitem{kn:albrecht}
Andreas Albrecht, Cosmic Inflation and the Arrow of Time, in
\emph{Science and Ultimate Reality: Quantum Theory, Cosmology and
Complexity}, eds J. D. Barrow, P.C.W. Davies, C.L. Harper, Cambridge
University Press (2004), \x astro-ph/0210527
\bibitem{kn:price}
Huw Price, Cosmology, Time's Arrow, and That Old Double Standard, in
\emph{Time's Arrows Today}, ed S. Savitt, Cambridge University Press
1994, \x gr-qc/9310022; The Thermodynamic Arrow: Puzzles and
Pseudo-puzzles, \x physics/0402040
\bibitem{kn:carroll}
Sean M. Carroll, Jennifer Chen, Spontaneous Inflation and the Origin
of the Arrow of Time, \x hep-th/0410270
\bibitem{kn:coule}
D.H. Coule, Quantum Cosmological Models, Class.Quant.Grav. 22 (2005)
R125, \x gr-qc/0412026
\bibitem{kn:dolgov}
A.D. Dolgov, Cosmology and New Physics, \x hep-ph/0606230
\bibitem{kn:trodden}
Tanmay Vachaspati, Mark Trodden, Causality and Cosmic Inflation,
Phys.Rev. D61 (2000) 023502, \x gr-qc/9811037
\bibitem{kn:mukhanov}
Lev Kofman, Andrei Linde, V. Mukhanov, Inflationary Theory and
Alternative Cosmology, JHEP 0210 (2002) 057, \x hep-th/0206088
\bibitem{kn:hollands}
Stefan Hollands, Robert M. Wald, Comment on Inflation and
Alternative Cosmology, \x hep-th/0210001
\bibitem{kn:juan}
Juan M. Maldacena, Eternal Black Holes in AdS, JHEP 0304 (2003) 021,
\x hep-th/0106112
\bibitem{kn:stephen}
Stephen Hawking, Information Loss in Black Holes, Phys.Rev. D72
(2005) 084013, \x hep-th/0507171
\bibitem{kn:sbound}
Brett McInnes, Unitarity at Infinity and Topological Holography,
Nucl.Phys. B754 (2006) 91, \x hep-th/0606068
\bibitem{kn:penrose}
R. Penrose, Singularities and Time-Asymmetry, in \emph{General
Relativity: An Einstein Centenary Survey}, eds S W Hawking, W
Israel, Cambridge University Press, 1979
\bibitem{kn:kobayashi}
S. Kobayashi, K. Nomizu {\em Foundations of Differential Geometry
I}, Interscience, 1963
\bibitem{kn:ooguri}
Hirosi Ooguri, Cumrun Vafa, Erik Verlinde, Hartle-Hawking
Wave-Function for Flux Compactifications: The Entropic Principle,
Lett.Math.Phys. 74 (2005) 311, \x hep-th/0502211
\bibitem{kn:OVV}
Brett McInnes, The Geometry of The Entropic Principle and the Shape
of the Universe, JHEP 10 (2006) 029, \x hep-th/0604150
\bibitem{kn:end}
John McGreevy, Eva Silverstein, The Tachyon at the End of the
Universe, JHEP 0508 (2005) 090, \x hep-th/0506130
\bibitem{kn:vilenkin}
A. Vilenkin, Creation of Universes from Nothing, Phys.Lett.B117
(1982) 25
\bibitem{kn:hartle}
J.B. Hartle and S.W. Hawking, Wave Function of the Universe,
Phys.Rev.D28 (1983) 2960
\bibitem{kn:laura}
R. Holman, L.Mersini-Houghton, Why did the Universe Start from a Low
Entropy State?, \x hep-th/0512070; Laura Mersini-Houghton, The Arrow
of Time Forbids a Positive Cosmological Constant $\Lambda$, \x
gr-qc/0609006
\bibitem{kn:gorsky}
A.Gorsky, Spontaneous Creation of the Brane World and Direction of
the Time Arrow, \x hep-th/0606072
\bibitem{kn:borde}
Arvind Borde, Alex Vilenkin, Phys.Rev. D56 (1997) 717, \x
gr-qc/9702019; Arvind Borde, Alan H. Guth, Alexander Vilenkin,
Phys.Rev.Lett. 90 (2003) 151301, \x gr-qc/0110012
\bibitem{kn:susskind}
Ben Freivogel, Matthew Kleban, Maria Rodriguez Martinez, Leonard
Susskind, Observational Consequences of a Landscape, JHEP 0603
(2006) 039, \x hep-th/0505232
\bibitem{kn:bat}
Thorsten Battefeld, Scott Watson, String Gas Cosmology,
Rev.Mod.Phys. 78 (2006) 435, \x hep-th/0510022
\bibitem{kn:osv}
Hirosi Ooguri, Andrew Strominger, Cumrun Vafa, Black Hole Attractors
and the Topological String, Phys.Rev. D70 (2004) 106007, \x
hep-th/0405146
\bibitem{kn:xi}
Chris Beasley, Davide Gaiotto, Monica Guica, Lisa Huang, Andrew
Strominger, Xi Yin, Why $\m{Z_{BH} = |Z_{top}|^2}$, \x
hep-th/0608021
\bibitem{kn:pioline}
Boris Pioline, Lectures on on Black Holes, Topological Strings and
Quantum Attractors, Class.Quant.Grav. 23 (2006) S981, \x
hep-th/0607227
\bibitem{kn:gall}
Gregory J. Galloway, Cosmological Spacetimes with $\Lambda\;>\;0$,
\x gr-qc/0407100
\bibitem{kn:zelda}
Ya.B.Zel'dovich, A.A. Starobinsky, Quantum Creation of a Universe
with Nontrivial Topology, Pis'ma Astron. Zh. 10 (1984) 323
\bibitem{kn:lindetypical}
Andrei Linde, Creation of a Compact Topologically Nontrivial
Inflationary Universe, JCAP 0410 (2004) 004, \x hep-th/0408164
\bibitem{kn:tallandthin}
Brett McInnes, The Most Probable Size of the Universe, Nucl.Phys.
B730 (2005) 50, \x hep-th/0509035
\bibitem{kn:andergall}
L. Andersson, G.J. Galloway, dS/CFT and spacetime topology,
Adv.Theor.Math.Phys. 6 (2003) 307, arXiv:hep-th/0202161
\bibitem{kn:singularstable}
Brett McInnes, Pre-Inflationary Spacetime in String Cosmology, Nucl.
Phys. B748 (2006) 309, \x hep-th/0511227
\bibitem{kn:crem}
Paolo Creminelli, Markus A. Luty, Alberto Nicolis, Leonardo
Senatore, Starting the Universe: Stable Violation of the Null Energy
Condition and Non-standard Cosmologies, JHEP 0612 (2006) 080, \x
hep-th/0606090
\bibitem{kn:ovrut}
Evgeny I. Buchbinder, Justin Khoury, Burt A. Ovrut, New Ekpyrotic
Cosmology, \x hep-th/0702154
\bibitem{kn:cremi}
Paolo Creminelli, Leonardo Senatore, A smooth bouncing cosmology
with scale invariant spectrum, \x hep-th/0702165
\bibitem{kn:nimah}
Nima Arkani-Hamed, Sergei Dubovsky, Alberto Nicolis, Enrico
Trincherini, Giovanni Villadoro, A Measure of de Sitter Entropy and
Eternal Inflation, arXiv:0704.1814
\bibitem{kn:god}
Wlodzimierz Godlowski, Marek Szydlowski, Towards observational
constraints on negative $\m{(1+z)^4}$ type contribution in the
Friedmann equation, Phys.Lett. B642 (2006) 13, \x  astro-ph/0606731
\bibitem{kn:kachru}
S. Kachru, L. McAllister, Bouncing Brane Cosmologies from Warped
String Compactifications, JHEP 0303 (2003) 018, arXiv:hep-th/0205209
\bibitem{kn:varun}
Varun Sahni, Cosmological Surprises from Braneworld models of Dark
Energy, \x astro-ph/0502032
\bibitem{kn:ruth}
Luis P. Chimento, Ruth Lazkoz, Roy Maartens, Israel Quiros, Crossing
the phantom divide without phantom matter, JCAP 09 (2006) 004, \x
astro-ph/0605450; Ruth Lazkoz, Roy Maartens, Elisabetta Majerotto,
Observational constraints on phantom-like braneworld cosmologies,
Phys.Rev. D74 (2006) 083510, \x astro-ph/0605701
\bibitem{kn:star}
Radouane Gannouji, David Polarski, Andre Ranquet, Alexei A.
Starobinsky, Scalar-Tensor Models of Normal and Phantom Dark Energy,
JCAP 0609 (2006) 016, \x astro-ph/0606287
\bibitem{kn:peri}
S. Nesseris, L. Perivolaropoulos, Crossing the Phantom Divide:
Theoretical Implications and Observational Status, JCAP 0701 (2007)
018 \x astro-ph/0610092
\bibitem{kn:nood}
Shin'ichi Nojiri, Sergei D. Odintsov, Modified gravity and its
reconstruction from the universe expansion history, \x
hep-th/0611071

\bibitem{kn:gab}
Gregory Gabadadze, Yanwen Shang, Classically Constrained Gauge
Fields and Gravity, \x hep-th/0506040; Gregory Gabadadze, Yanwen
Shang, Quantum Cosmology of Classically Constrained Gravity,
Phys.Lett. B635 (2006) 235, \x hep-th/0511137
\bibitem{kn:stach}
Tomasz Stachowiak, Marek Szydlowski, Exact solutions in bouncing
cosmology, Phys.Lett. B646 (2007) 209, \x gr-qc/0610121
\bibitem{kn:where}
Shin'ichi Nojiri, Sergei D. Odintsov, Where new gravitational
physics comes from: M-theory?, Phys.Lett. B576 (2003) 5, \x
hep-th/0307071
\bibitem{kn:biswas}
Tirthabir Biswas, Anupam Mazumdar, Warren Siegel, Bouncing Universes
in String-inspired Gravity, JCAP 0603 (2006) 009, \x hep-th/0508194;
Tirthabir Biswas, Robert Brandenberger, Anupam Mazumdar, Warren
Siegel, Non-perturbative Gravity, Hagedorn Bounce \& CMB, \x
hep-th/0610274
\bibitem{kn:ghost}
S. Nojiri, S.D. Odintsov, Introduction to Modified Gravity and
Gravitational Alternative for Dark Energy, Int.J.Geom.Meth.Mod.Phys.
4 (2007) 115, hep-th/0601213
\bibitem{kn:syd}
Marek Szydlowski, Cosmological zoo -- accelerating models with dark
energy, \x astro-ph/0610250
\bibitem{kn:maoz}
Juan Maldacena, Liat Maoz, Wormholes in AdS, JHEP 0402 (2004) 053,
\x hep-th/0401024
\bibitem{kn:hawking}
S.W. Hawking , The Arrow Of Time In Cosmology, Phys.Rev.D32 (1985)
2489
\bibitem{kn:laf}
S.W. Hawking, R. Laflamme, G.W. Lyons, The Origin of time asymmetry,
Phys.Rev.D47 (1993) 5342 \x gr-qc/9301017
\bibitem{kn:gibhart}
G.W. Gibbons, J.B. Hartle, Real Tunneling Geometries and the
Large-Scale Topology of the Universe, Phys.Rev. D42 (1990) 2458
\bibitem{kn:maldacena}
Brett McInnes, Quintessential Maldacena-Maoz Cosmologies, JHEP 0404
(2004) 036, \x hep-th/0403104
\bibitem{kn:zeh}
C. Kiefer, H.D. Zeh, Arrow of time in a recollapsing quantum
universe, Phys.Rev.D51 (1995) 4145 \x gr-qc/9402036
\bibitem{kn:dab1}
M.P. Dabrowski, A.L. Larsen, Quantum Tunneling Effect in Oscillating
Friedmann Cosmology, Phys.Rev. D52 (1995) 3424, \x gr-qc/9504025
\bibitem{kn:dab2}
Mariusz P. Dabrowski, Claus Kiefer, Barbara Sandhoefer, Quantum
phantom cosmology, Phys.Rev. D74 (2006) 044022, \x hep-th/0605229
\bibitem{kn:twamley}
P C W Davies and J. Twamley, Time-symmetric cosmology and the
opacity of the future light cone, Class. Quantum Grav. 10 (1993) 931
\bibitem{kn:vafa}
Sergei Gukov, Kirill Saraikin, Cumrun Vafa, The Entropic Principle
and Asymptotic Freedom, Phys.Rev. D73 (2006) 066010, \x
hep-th/0509109
\bibitem{kn:wald}
Robert M. Wald, The Arrow of Time and the Initial Conditions of the
Universe, \x gr-qc/0507094
\bibitem{kn:waldbook}
Robert M. Wald, \emph{General Relativity}, Chicago University Press,
1984
\bibitem{kn:milnor}
John W. Milnor, James D. Stasheff, \emph{Characteristic Classes},
Princeton University Press, 1974
\bibitem{kn:gell} Murray Gell-Mann,
James Hartle, Quasiclassical Coarse Graining and Thermodynamic
Entropy, \x quant-ph/0609190
\bibitem{kn:porrati}
M. Kleban, M. Porrati, R. Rabadan, Stability in Asymptotically AdS
Spaces, JHEP 0508 (2005) 016, \x hep-th/0409242
\bibitem{kn:kazdan}
Kazdan, Jerry L., Warner, F. W., Existence and conformal deformation
of metrics with prescribed Gaussian and scalar curvatures. Ann. of
Math. 101 (1975) 317
\bibitem{kn:schoenyau}
R. Schoen, S.-T. Yau, Existence of incompressible minimal surfaces
and the topology of three-dimensional manifolds with nonnegative
scalar curvature, Ann. of Math. 110 (1979) 127
\bibitem{kn:lawson}
H. Blaine Lawson and Marie-Louise Michelsohn, \emph{Spin Geometry},
Princeton University Press, 1990
\bibitem{kn:perelman}
Grisha Perelman, Finite extinction time for the solutions to the
Ricci flow on certain three-manifolds, \x math.DG/0307245
\bibitem{kn:milnor2}
J. Milnor, A unique decomposition theorem for 3-manifolds, Amer. J.
Math. 84 (1962) 1
\bibitem{kn:wolf}
J.A. Wolf, \emph{Spaces of Constant Curvature}, Fifth Edition,
Publish or Perish Press, 1984
\bibitem{kn:weeks}
Roland Lehoucq, Jeffrey Weeks, Jean-Philippe Uzan, Evelise Gausmann,
Jean-Pierre Luminet, Eigenmodes of 3-dimensional spherical spaces
and their application to cosmology, Class.Quant.Grav. 19 (2002)
4683, \x gr-qc/0205009
\bibitem{kn:schoen}
R. Schoen, S.-T. Yau, On the Structure of Manifolds with Positive
Scalar Curvature, Manuscripta Math. 28 (1979) 159
\bibitem{kn:conway}
John Horton Conway, Juan Pablo Rossetti, Describing the platycosms,
\x math.DG/0311476
\bibitem{kn:aiello}
Matias Aiello, Mario Castagnino, Olimpia Lombardi, The arrow of
time: from universe time-asymmetry to local irreversible processes,
\x gr-qc/0608099

\bibitem{kn:mixing}
Neil J. Cornish, David N. Spergel, Glenn D. Starkman, Does Chaotic
Mixing Facilitate $\Omega<1$ Inflation?, Phys.Rev.Lett. 77 (1996)
215, \x astro-ph/9601034

\bibitem{kn:polchinski}
Joseph Polchinski, The Cosmological Constant and the String
Landscape, \x hep-th/0603249
\bibitem{kn:bruhat}
Yvonne Choquet-Bruhat, James Isenberg, Daniel Pollack, Applications
of theorems of Jean Leray to the Einstein-scalar field equations, \x
gr-qc/0611009
\bibitem{kn:topologically}
Brett McInnes, Topologically Induced Instability in String Theory,
JHEP 0103 (2001) 031, \x hep-th/0101136


\bibitem{kn:page}
Don N. Page, Is Our Universe Decaying at an Astronomical Rate?, \x
hep-th/0612137
\bibitem{kn:carlip}
S. Carlip, Transient Observers and Variable Constants, or Repelling
the Invasion of the Boltzmann's Brains, \x hep-th/0703115
\bibitem{kn:sred}
James B. Hartle, Mark Srednicki, Are We Typical?, arXiv:0704.2630

\bibitem{kn:bousso}
Raphael Bousso, Precision Cosmology and the Landscape, \x
hep-th/0610211
\end{thebibliography}
\end{document}